# Partially coherent double phase holography in visible using meta-optics

Saswata Mukherjee[1,*], Quentin A. A. Tanguy[1], Johannes E. Fröch[1], Aamod Shanker[1], Karl F. Böhringer[1,3], Steven Brunton[4] and Arka Majumdar[1,2,*]

[1]Department of Electrical and Computer Engineering, University of Washington, Seattle, WA, USA

[2]Department of Physics, University of Washington, Seattle, WA, USA

[3]Institute for Nano-engineered Systems, University of Washington, Seattle, WA, USA

[4]Department of Mechanical Engineering, University of Washington, Seattle, WA, USA

* Corresponding author: arka@uw.edu, saswata@uw.edu

## Abstract

Ultrathin flat meta-optics have shown great promise for holography in recent years. However, most of the reported meta-optical holograms rely on only phase modulation and neglect the amplitude information. Modulation of both amplitude and phase in meta-optics either requires polarization sensitive meta-atoms, or complex scatterers with stringent fabrication requirements. Additionally, almost all the meta-optical holograms were measured under laser illumination. Here we adopt the concept of double-phase holography, to report polarization-independent holography with both amplitude and phase modulation, using dielectric meta-optics. We validate the implementation of complex phase hologram by measuring an improvement of structural similarity of the reconstructed hologram by ∼3 × over phase-only holograms. Finally, we demonstrate that meta-optical holography can also be realized using partially incoherent light from a light emitting diode. This observation can significantly reduce the alignment complexity and speckles in laser-based meta-optical holography.

## Introduction

Holography is a key concept for next generation display technologies, widely being explored for three-dimensional (3D) displays and near eye visors for augmented and virtual reality, where an immersive visual experience for the user is of utmost importance. The fundamental principle of holography relies on the reconstruction of the light field produced by a screen containing both amplitude and phase information[1]. The simplest configuration, to store and recreate a hologram is achieved by recording the interference patterns formed by light scattered by an object and a reference coherent beam[1]. While this technique creates crisp scenes, the requirement of recording a real object and a spatially and temporally coherent light source severely restricts applications, as typically only existing and static scenes are recordable. These particular limitations were resolved with the advent of digital computers and the emergence of spatial light modulators (SLM), which brought forth the development of computer generated holography (CGH)[2,3]. In CGH, the phase profile for any desired image is generated via computational algorithms, and subsequently the phase profile is displayed on an SLM, which under coherent light illumination creates the image at the desired plane. Based on the same principle, static holograms can be implemented using patterned surfaces, such as diffractive optics[4]. However, traditional SLMs or diffractive optics cannot modulate both amplitude and phase, making CGHs primarily limited to either phase-only or amplitude-only modulation. This typically results in a poor signal-to-noise ratio (SNR) of the reconstructed image. A solution to this problem is achieved by implementing both amplitude and phase modulation via dual phase holography [5–7], where each value of the complex phase is coded using two different pixels. While dual phase holography improves the SNR, it comes at the cost of reduced spatial resolution[8].

In recent years, meta-optics has emerged as a promising alternative to implement a desired phase profile. Meta-optics are artificially manufactured arrays of sub-wavelength scatterers [9,10], which shape the optical wavefront with high spatial resolution. A distinct advantage is that the sub-wavelength pitch of scatterers precludes any higher-order diffraction, thus maximizing light in the zero-th order diffraction and making it more efficient than traditional diffractive optics [11,12]. These properties have inspired many researchers to demonstrate holography with meta-optics [13–18]. However, for the majority of these works, phase-only modulation has been considered. To implement complex phase holography, i.e. both phase and amplitude modulation, meta-atoms with complex orientations and shapes were utilized in THz domain [19,20], but their fabrication will be difficult in the visible wavelength. The other approaches are generally polarization sensitive [21,22]. Complex phase modulation can also be achieved using multi-layer meta-optics[23,24], where the first meta-optic modulates the amplitude and a subsequent one modulates the phase of the transmitted light. However, such bilayer meta-optics are arduous to fabricate due to their requirement of precise alignments. Additionally, one needs to rely on high-contrast materials like silicon or metal as they need to be embedded in polymers. This presents a limitation for low-index materials like silicon nitride[25] of titanium dioxide[26], which are transparent in the visible.

In this work, we report polarization-independent complex phase modulated holography using a single visible meta-optics containing simple square pillar meta-atoms. We employ double phase holography to implement the complex phase modulation. Thanks to he sub-wavelength pitch in a meta-optic, we can achieve very high space-bandwidth product and thus the main limitation of double phase holography is not as severe as in for meta-optic. We establish the efficacy of double-phase holography by demonstrating a higher structural similarity index metric (SSIM) of double-phase holograms over phase-only holograms, under both coherent and incoherent green illumination. The ability to create meta-optical holograms using incoherent illumination from light emitting diodes can significantly simplify the optical arrangement making them more applicable for practical applications, including near eye displays for augmented and virtual reality [27]. Additionally, the ability to coherently manipulate the optical field will be important for free-space vector-matrix multiplication and optical computing [28].

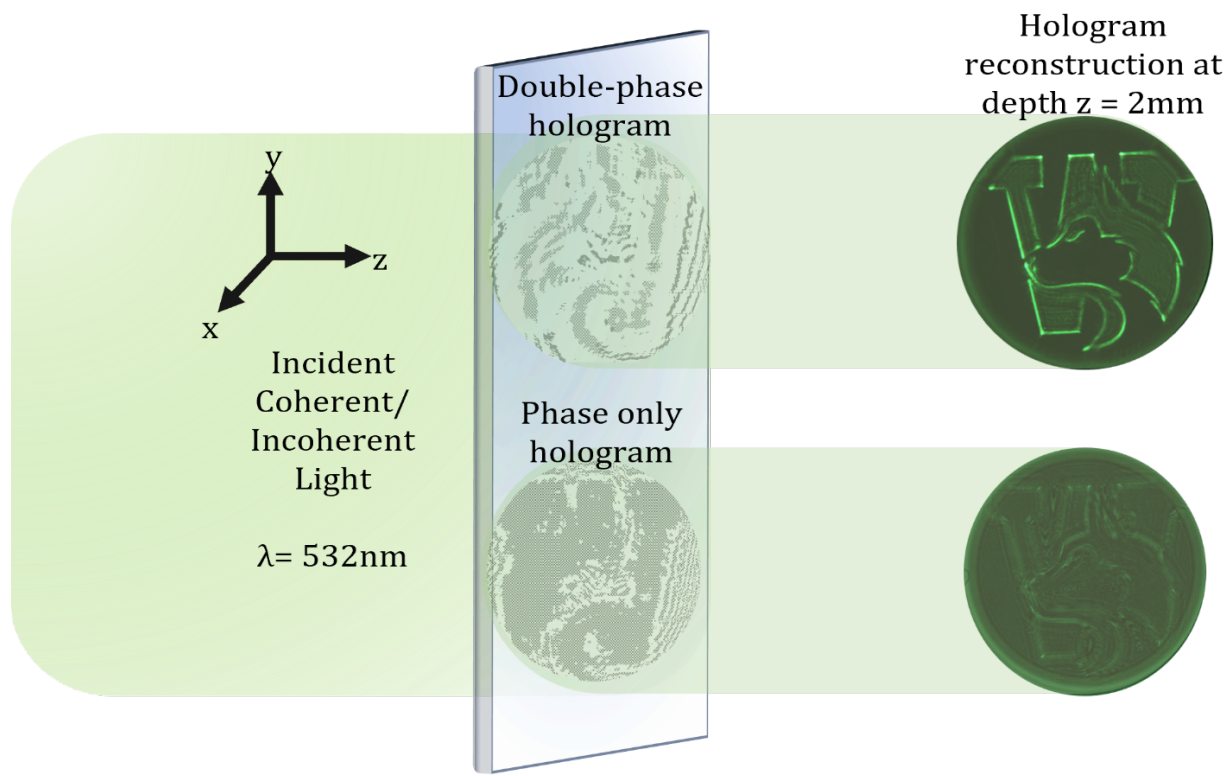

*Figure 1: Illustration of holography using a double phase hologram and a phase-only hologram realized by all-dielectric meta-optics. Under incident green light (both coherent and incoherent), the device displays an image at the desired plane. The image quality with double phase holography is expected to be better than simple phase-only hologram.*

## Methodology

Figure 1 shows the schematic where the meta-optical hologram is irradiated with a coherent or incoherent light source ($\lambda = 532$ nm), and the images are reconstructed at the imaging plane z=2 mm away from the meta-optic. The meta-optics is polarization independent and designed to generate the images at the desired wavelength. Here we used two objects, "digit W and Husky," and a part of the United States Air Force (USAF) test targets. We assumed the object to be virtually located at the imaging plane (z = 2mm) from the meta-optics and used the angular spectrum method [29] to estimate the desired complex phase of the meta-optic. We applied the double phase encoding technique to generate the complex field in the meta-hologram. We also extracted the real part of the phase from the complex electric field to realize a phase-only meta-optic hologram. Additionally, another phase-only meta-optic hologram was realized using the Gerchberg-Saxton algorithm [30], having imaging plane and size identical to other holograms.

The procedure for realizing the double phase hologram is shown in Figure 2. Let us assume the target complex hologram $U_F$ can be expressed as $U_F(x,y) = A_1(x,y) \cdot e^{i\varphi(x,y)}$, where $\varphi(x,y)$ is the phase and $A_1(x,y)$ is normalized amplitude with values in the range between 0 and 2. We can then rewrite $A_1(x,y) = e^{i\theta(x,y)} + e^{-i\theta(x,y)} = 2\cos(\theta(x,y))$, where $\theta(x,y) \in [-\pi, \pi]$. Hence, we can rewrite $U_F$ as

$$U_F(x,y) = h_1(x,y) \times M_1 + h_2(x,y) \times M_2$$

with $h_1 = e^{i[\varphi(x,y)+\theta(x,y)]}$ and $h_2 = e^{i[\varphi(x,y)-\theta(x,y)]}$. Thus, at each point $(x,y)$, we obtain two phase values: $\phi(x,y) \pm \theta(x,y)$. The complementary binary masks ($M_1$ and $M_2$)[31] with 2D checkerboard patterns, helps reduce the required phase at each point to a single value.

This sampling technique is similar to the random mask encoding for multiplexing phase-only filters[32], except we select two complementary binary functions under Nyquist limit. This phase profile can then be translated into a meta-optic and upon illumination with light, a hologram forms at the desired plane, as schematically depicted in Figure 1.

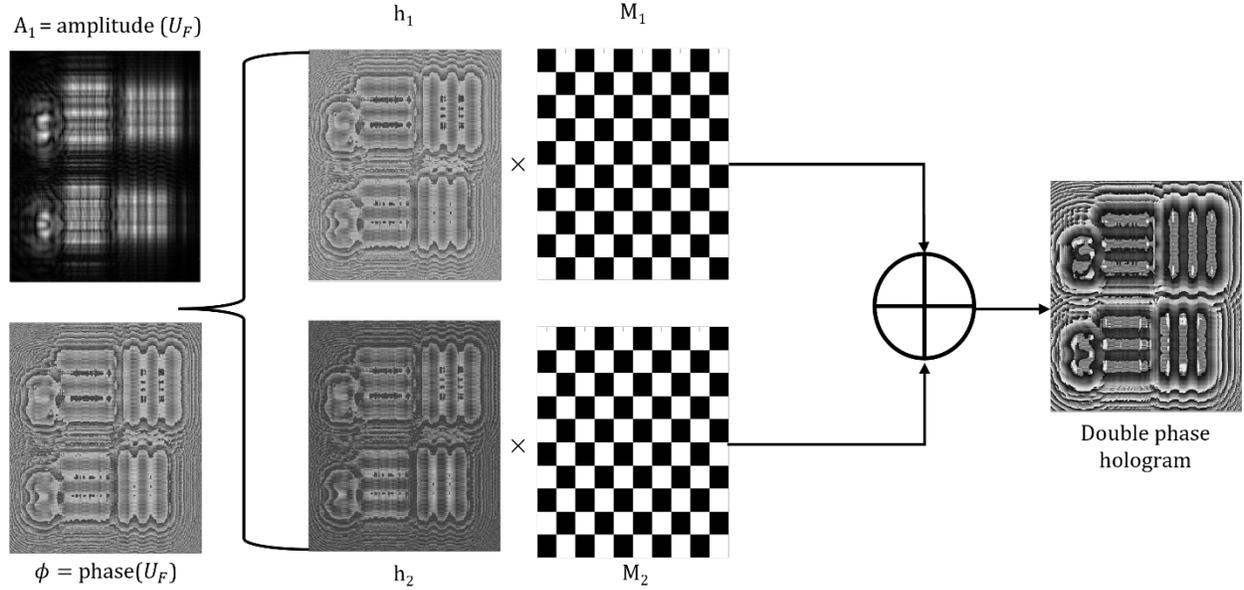

*Figure 2:* Flowchart of double phase encoding method for complex field modulation. a) First the complex hologram $U_F$ is calculated from the desired ground truth. Then the amplitude and phase from the complex field is extracted for decompositions of the given complex field into two pure phase functions $h_1$ and $h_2$, using the phase encoding technique[5]. Two complementary 2D binary masks (checkerboard patterns) are superimposed on the two pure phase elements to reduce it to one phase value at every coordinate (x,y).

## Design

The meta-optical hologram consists of an array (periodicity $p = 350$ nm) of meta-atoms, here sub-wavelength silicon nitride square pillars with height $t = 530$ nm on quartz substrate (Figure 3a). We keep a blanket 70nm-thick silicon nitride film on the quartz substrate to ensure mechanical stability. The width $w$ of the pillars are varied between 75 nm and 330 nm to cover the entire $2\pi$ phase modulation at a wavelength of $\lambda = 530\ nm$ with unity transmission (Figure 3b), as simulated using rigorous coupled-wave analysis (RCWA) [33]. These meta-atoms also ensure a moderate aspect ratio and can be easily fabricated. To design the meta-optics, we selected 12 different scatterers (all with transmission greater than 85%), i.e., the resulting hologram has 12 discrete phase levels. Figure 3c shows the scanning electron microscope images of the fabricated meta-optic (details in Methods).

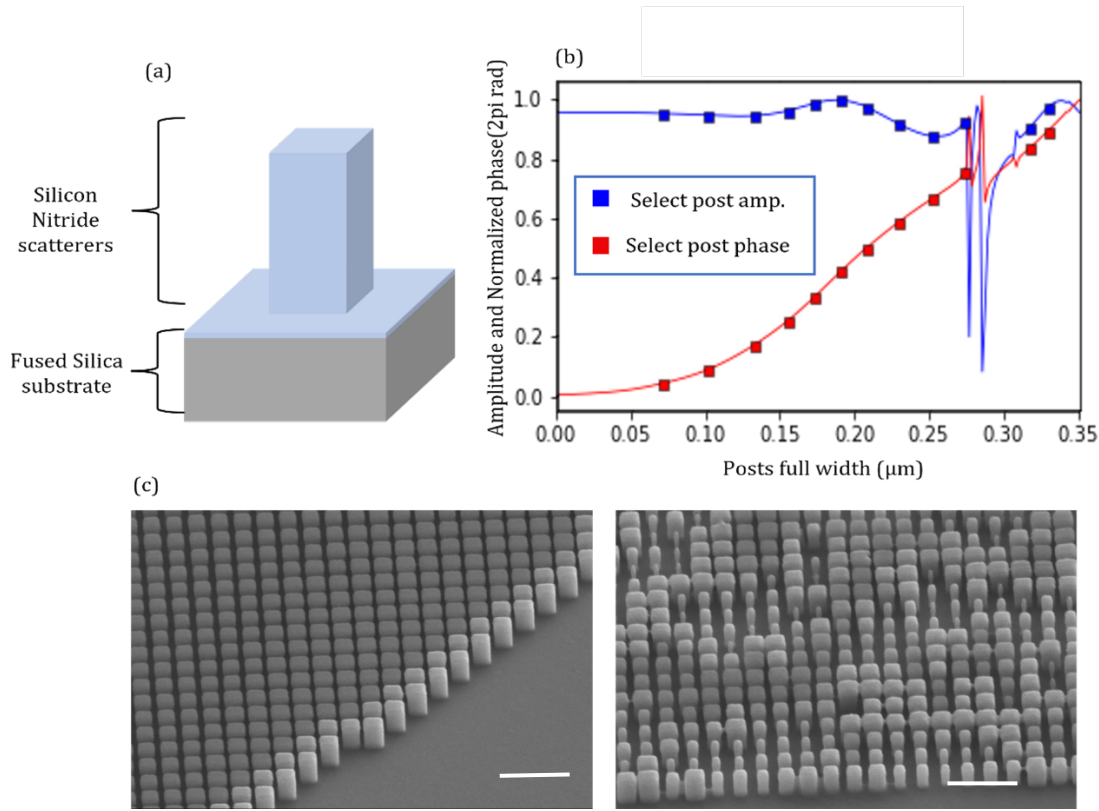

*Figure 3:* *(a) Schematic of silicon nitride pillars with width w, pitch p and height t. (b) The transmission amplitude and phase of the nano-pillars as a function of the pillar width, simulated using RCWA[33]. (c) Scanning electron microscope (SEM) images of part of the meta-optical hologram showing the nano-pillars. Scalebar: 1μm.*

## Results

The efficacy of the double phase holograms over the phase-only holograms was then experimentally assessed. Either a laser or a light emitting diode is used to illuminate the meta-holograms (Figure 4a), and the reconstructed image was captured using a movable microscope (details in Methods).

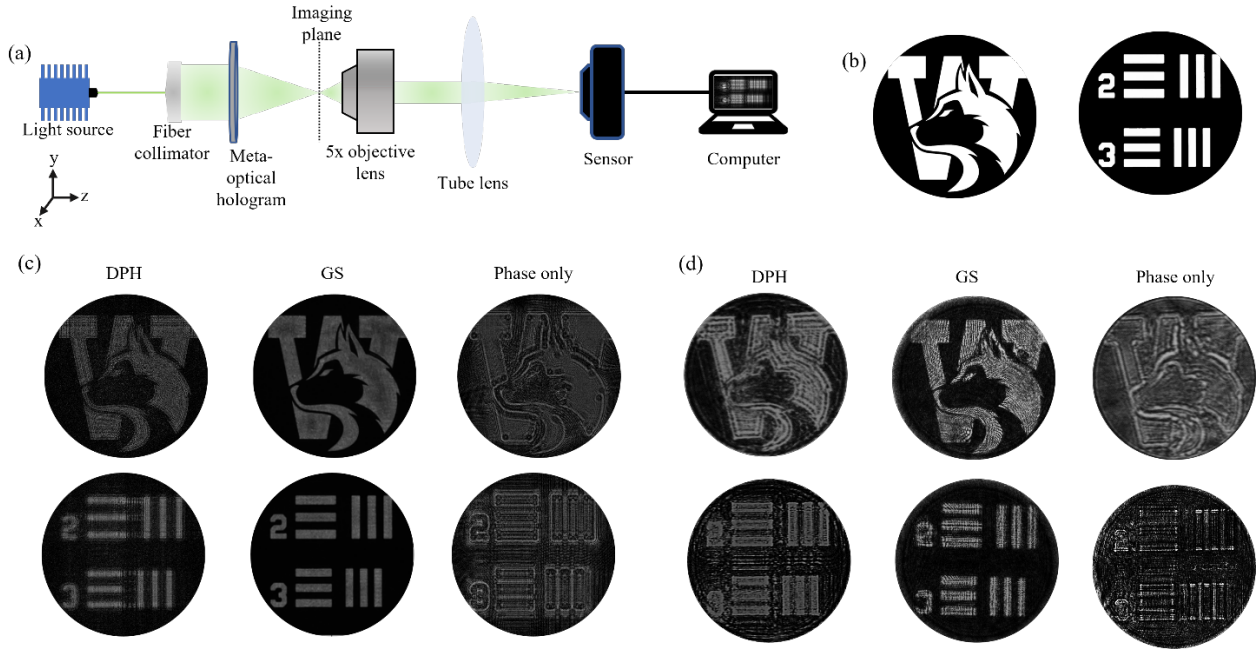

*Figure 4:* a) Schematic of optical setup to test the meta-optical holography. b) The appropriately cropped original object patterns of "W and Husky" and "USAF target portion" are the ground truth image. c) Simulated and d) Experimentally measured holographic reconstruction for three methods: double phase holography (DPH), Gerchberg-Saxton holography (GS) and amplitude discarded phase-only holography,

Using two different ground truth images (Fig. 4b), we first compare the performance of the holography for three cases under coherent green illumination: complex phase with double phase holography, phase-only hologram with only real part of the complex phase, and phase-only hologram designed via Gerchberg-Saxton algorithm. A low pass filter was applied to the target image in order to blur the boundaries of the object so that the outer edges of the hologram have low amplitudes. This correctly modulates the higher spatial frequencies in the hologram and removes the unwanted edge enhancements (see Supplementary Information). Both in the simulation (Fig. 4c) and experiment (Fig. 4d), the double phase holography shows better reconstruction results compared to phase-only holography and comparable results to GS holograms. However, the GS holograms have the problems of unwanted speckles in reconstruction due to destructive interference occurring within the target image region. To quantitatively validate our claim, we calculated the structural similarity metrics (SSIM) between the reconstructed images and the ground truth in the region of interest containing only the image (Table-1). We clearly observe superior performance of the double phase holography.

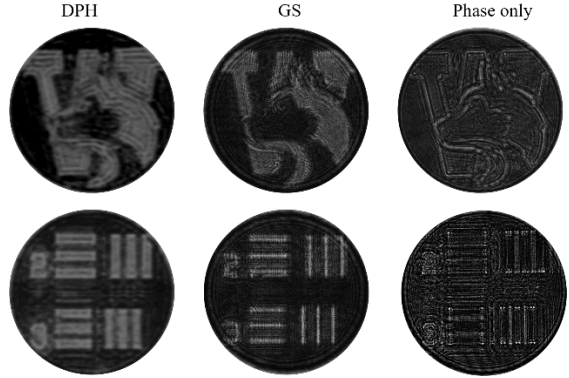

*Figure 5: Hologram reconstruction with incoherent illumination using green light emitting diodes with a bandwidth of ~20nm. DPH- Double phase hologram; GS- Gerchberg-Saxton hologram; Phase-only holograms*

Although for holography coherent illumination typically yields better results, incoherent light sources such as light emitting diodes can be significantly more cost effective. Additionally, the speckle under incoherent illumination is minimal, whereas under laser illumination we often need to rely on extra optics to reduce spectral coherence. Hence, we tested the meta-optical holograms under incoherent illumination. Here we use a green light emitting diode (Thorlabs M530F1) at a central wavelength of 525 nm and linewidth ~20 nm. The resulting holograms are shown in Figure 5. While coherent illumination does provide better holographic reconstruction, we demonstrate that a comparable quality for reconstruction can be achieved using incoherent illumination, as can be seen in the SSIMs in Table 1. Furthermore, we observed that the double phase holograms achieved higher SSIM even under incoherent illumination compared to the amplitude discarded phase-only approach under coherent illumination. We emphasize that, this is the first-time incoherent holography has been reported using meta-optics, which can potentially enable a drastic reduction of the form factor of near-eye visors[34].

**Table 1:** Structural similarity metrics of different types of measured holograms for two different objects using both coherent and incoherent illumination.

|  | DPH | | GS | | Phase-only | |
| --- | --- | --- | --- | --- | --- | --- |
|  | Coherent | Incoherent | Coherent | Incoherent | Coherent | Incoherent |
| **USAF target portion** | 0.5834 | 0.5703 | 0.6408 | 0.5144 | 0.2621 | 0.2185 |
| **W and Husky** | 0.5855 | 0.5496 | 0.5728 | 0.5251 | 0.2736 | 0.2301 |

# Discussion

We have demonstrated complex field modulated holography in meta-optics using a double phase encoding technique. Our proposed approach is polarization independent and does not rely on complicated meta-atoms. Moreover, we demonstrated meta-optical holography under an incoherent light emitting diode source. While partially incoherent holography has been reported in CGH literature in the past, this is the first time it is demonstrated in a meta-optic. We compared three meta-hologram types (complex phase, amplitude-discarded phase-only and phase-only via GS algorithm) using SSIM and demonstrated superiority of double phase hologram. This work offers a robust and generalizable method for realizing the primary promise of meta-optics: to modulate complex field at will. Such ability to modulate complex field will be beneficial for other applications, including free-space optical computing.

# Acknowledgement

The research is supported by NSF- 2040527. Part of this work was conducted at the Washington Nanofabrication Facility / Molecular Analysis Facility, a National Nanotechnology Coordinated Infrastructure (NNCI) site at the University of Washington with partial support from the National Science Foundation via awards NNCI-1542101 and NNCI-2025489.

# Materials and Methods

### Fabrication process

We fabricated the meta-optical holograms using a 500-micron thick fused silica wafer and depositing 600 nm of silicon nitride using plasma enhanced chemical vapor deposition at 350-degree centigrade. A 300 nm thick layer of ZEP 520A followed by a thin film of anti-charging polymer (DisCharge $H_2O$) was spun coat on top of $Si_3N_4$ thin film. Next, the hologram patterns are written by electron beam lithography (JEOL 6300) at a beam voltage of 100 kV, beam current of 8000 pA, and a base dose of 275 $\mu C/cm^2$ and appropriate proximity effect corrections. The resulting designs are developed in a solution of amyl acetate and cleaned with Iso-Propyl alcohol. The exposed and developed samples are then placed in a physical evaporator to deposit roughly 60 nm of aluminum oxide. The dissolution of the remaining resist performs lift-off in N-

Methyl-2-pyrrolidone (NMP) at 90°C for 12 hours. Finally, the pattern is transferred from the aluminum oxide mask to the $Si_3N_4$ by using a fluorine based RIE process (Oxford) leaving a total thickness of 10 nm of Alumina over 530 nm of $Si_3N_4$.

**Optical Characterization**

A set of collimating optics passes coherent light (Laserglow; wavelength: 532nm; max power :150 mw) to the meta-optics. Light is collected and analyzed using a movable microscope. An infinity-corrected 5× objective (Nikon Plan Fluor 5X, 0.15 NA) collects light scattered by the meta-optic at 2mm away from the optic and passes it through a tube lens (Thorlabs, f=200 mm. Then an iris is used to cut off unwanted light from reaching the CMOS sensor (Point Grey CMLN 13S2M CS). The same experiment was repeated with incoherent light by replacing the green laser with the green light emitting diode (Thorlabs M530F1), while keeping the rest of the setup intact.

# Supporting Information

# Partially coherent double phase holography in visible using meta-optics

*Saswata Mukherjee[1,*], Quentin A. A. Tanguy[1], Johannes E. Fröch[1], Aamod Shanker[1], Karl F. Böhringer[1,3], Steven Brunton[4] and Arka Majumdar[1,2,*]*

*[1]Department of Electrical and Computer Engineering, University of Washington, Seattle, WA, USA*

*[2]Department of Physics, University of Washington, Seattle, WA, USA*

*[3]Institute for Nano-engineered Systems, University of Washington, Seattle, WA, USA*

*[4]Department of Mechanical Engineering, University of Washington, Seattle, WA, USA*

*\* Corresponding author: [arka@uw.edu](arka@uw.edu), [saswata@uw.edu](saswata@uw.edu)*

**Section S1. Effect of blur on reconstruction**

The free-space momentum of light is a fixed value constrained by the wavelength of light, and there is an upper limit to the spatial frequencies encodable by a meta-optics hologram. Simple concepts from Fourier analysis predict that a hologram cannot produce perfectly sharp boundaries in a 2D holographic image because arbitrarily large spatial frequencies encode this boundary. Attempting to reconstruct a perfectly sharp boundary with a finite range of spatial frequencies results in apparent amplitude ripples near the sharp boundary. To improve the aesthetic considerations, perfectly sharp boundaries should be smoothed out so that the experiment can faithfully encode the entire range of spatial frequencies represented by the holographic object. For our implementation, we apply a Gaussian blur to a target image (such as the "digit W and Husky" and a part of the United States Air Force (USAF) test targets) to eliminate the artifacts. The reconstruction with and without Low pass filtering is shown in Figure S1.

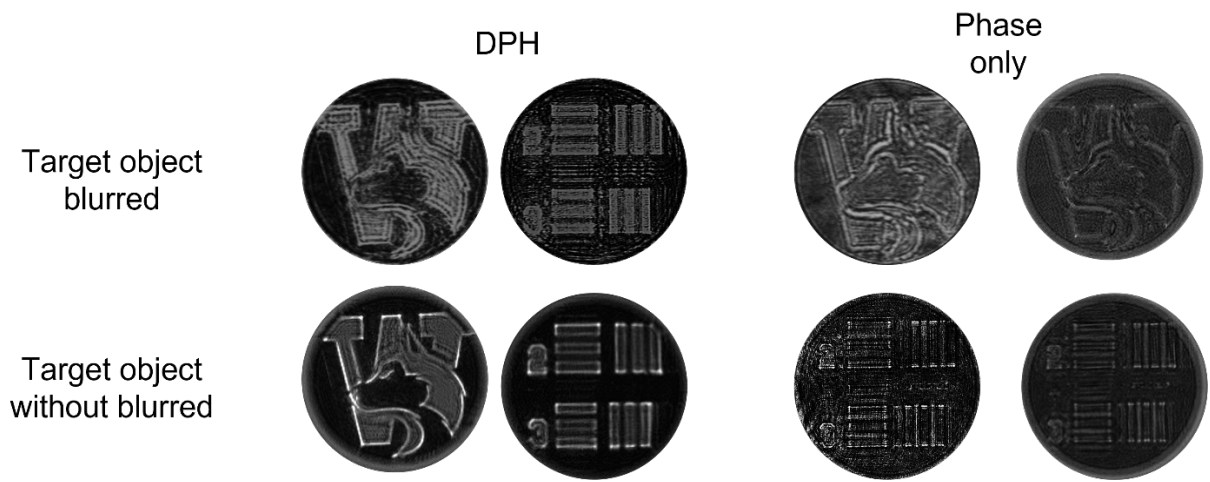

**Figure S1:** Hologram reconstruction with coherent illumination using green laser with and without the target object being blurred showing the effects of Higher spatial frequencies on boundaries of the objects. DPH- Double phase hologram; Phase-only holograms